\documentclass[a4paper,10pt,twoside]{article}
\usepackage{amsmath,amsfonts,amssymb,amsthm}
\usepackage{graphicx}
\pdfoutput=1
\usepackage{grffile}

\newtheorem{proposition}{Proposition}

\title{The dynamics of two-stage contagion}
\author{Guy Katriel\\ Department of Mathematics, ORT Braude College,\\ Karmiel, Israel\\}

\date{}

\begin{document}

\maketitle

\begin{abstract}
	We explore simple models aimed at the study of social contagion, in which contagion proceeds through two stages. When coupled with demographic turnover, we show that two-stage contagion leads to nonlinear phenomena which are not present in the basic `classical' models of mathematical epidemiology. These include: bistability, critical transitions, endogenous oscillations, and excitability,  suggesting that contagion models with stages could account for some aspects of the complex dynamics encountered in social life. These phenomena, and the bifurcations involved, are studied by a combination of analytical and numerical means. 
\end{abstract}

\section{Introduction}

The notion of social contagion has been gaining increasing prominence, with accumulating empirical
evidence of its importance for many aspects of our lives, from political mobilization, spread of ideas and innovations, and psychological well-being, to substance abuse, crime and violence, obesity, financial panics, and mass psychogenic illness \cite{centola2,christakis2,hatfield,lehman}.

Contagion phenomena have traditionally been mathematically modelled in the field of infectious disease epidemiology \cite{keeling,martcheva},
and it is natural to apply the tools developed in this field to social contagion, as is indeed being done by   
researchers from diverse fields \cite{castellano,galam,kribs,raafat,sooknanan}. Models for various instances of 
social contagion have been developed, including smoking \cite{castillo-garsow}, drug use \cite{white}, bulimia \cite{gonzalez}, political activism \cite{jeffs,mistry}, spread of rumors \cite{dietz}, crime \cite{dorsonga}, language dynamics \cite{loreto}, organized religions \cite{hayward}, diffusion of new products and technologies \cite{bass,fibich} or ideas in a scientific community \cite{bettencourt}, among many more examples.

It is important to address the various aspects in which social contagions differ from biological contagions at the level
of the individual, and the consequences of these differences for the dynamics of these phenomena at the population level.
In elucidating this micro-macro link, mathematical modelling plays an important role, as it allows us to study
the population-level patterns emerging from different contagion mechanisms, which are often far from intuitively obvious, 
and which may be quite different from those familiar from the study of classical models for the transmission of infectious diseases.

One important way in which social contagion mechanisms differ from those of biological contagion is that while infection with a 
pathogen is a discrete event ocurring upon contact of a susceptible individual with an infectious individual,
social contagion may require transitions through several stages, with each such transition dependent on contact with `infectives'.  This is a central tenet of Rogers' Diffusion of Innovations theory \cite{rogers}, which studies the spread of innovations - including, for example,
technologies, products, ideas, cultural practices, health-related behaviors, and more. Rogers' theory posits that the adoption of an innovation, at the individual level, 
involves a series of five stages in which the individual comes to learn about the innovation, assess it, and 
finally adopt it. The transition rate from one stage to another is modulated by the interpersonal influence of other individuals in one's social network, and hence
depends on the number of people who have already adopted the innovation. As another example from the social sciences, 
Klandermans and Oegema \cite{klandermans} analyse the process of becoming a participant in a social movement as consisting 
of four stages: becoming part of the mobilization potential, becoming the target of mobilization attempts, becoming motivated to 
participate, and overcoming barriers to participation. A closely related idea is that of {\it{complex contagion}} \cite{centola2},
referring to social contagions which require sustained or repeated contacts with adopters in order to spread.

To incorporate the idea of stages of contagion into mathematical models, one can divide the population into classes, each of which consists of individuals at a certain stage in the adoption process, with movement among the classes due to contact with adopters. Models of this type have been proposed studied in several works \cite{bizhani,choi1,choi2,chung,coletti,hasegawa,janssen1,janssen2,krapivski}, and in section \ref{previous} we will briefly describe  and
compare these with the model studied here.

Our aim here is to make a detailed study of the dynamics of a two-stage contagion model, which, unlike in nearly all previous works, incorporates the process of demographic turnover - recruitment and departure of individuals from the population. This may be due to births and deaths or, if considering a contagion spreading in particular institutional settings or age groups, individuals entering and leaving the institution, or maturing into and out of 
the relevant age group. Our model is thus a two-stage analog of the classical SIR model with demographic turnover \cite{keeling,martcheva}.
We will show that stages of contagion, in conjunction with demographic turnover, lead to new and sometimes surprising dynamical phenomena, which are {\it{not}} present in the basic one-stage models familiar in mathematical epidemiology.
The interesting behaviors we observe in our simple model suggest a possible generative mechanism for 
some of the complex phenomena observed in the social world: alternative stable states, discontinuous transitions, critical mass effects, and
periodic cycles. 

Since we wish to highlight the fact that stages of contagion lead to novel phenomena in the simplest of models,
we resist the temptation to generalize by incorporating more mechanisms or more stages of contagion.
The simplicity of the present models also provides the advantage that we 
can go quite far in characterizing their dynamics analytically, in different parameter regimes, using standard tools of stability analysis of equilibria. Some aspects of the dynamics, however, will be studied numerically, and obtaining full
mathematical proofs for some of our conclusions from these simulations seems like a challenging task for the future.
Any phenomena present in our minimal model will also occur {\it{a fortiori}} in more elaborate models, for some 
parameter values, and such models might give rise to further interesting dynamics not present in the two-stage model, and deserve to be studied further.

In the remainder of this section we introduce the two models to be studied, one in which an individual's adoption of an innovation is permanent 
and one in which it is temporary, and make a comparison with previous models incorporating two-stage contagion. The permanent adoption model will be studied in section \ref{permanent}, and the temporary
adoption model, which leads to a richer variety of possible dynamical behaviors, will be studied in section \ref{temporary}.
In section \ref{discussion} we will recap the novel features that the two-stage contagion models have in comparison with their
classical one-stage counterparts, and discuss their significance.

\subsection{The models}
\label{themodels}

We assume that the population is divided into three classes: class $S_0$ consists of `naive' individuals who 
have not been exposed to the innovation, class $S_1$ consists of `informed' individuals who have encountered the innovation 
but have not yet adopted it, and class $A$ consists of adopters. We choose the terminology of `adoption of innovations' for convenience;
the models could just as well describe potential supporters of a social movement ($S_0$), actual supporters ($S_1$), and activists ($A$), or 
many other examples of social contagion.

The mechanisms involved in the models are:
\begin{itemize}
	\item Adopters randomly encounter other individuals, at a 
	rate of $\beta$ effective contacts per unit time. 
	
	\item Upon effective contact with an adopter, a naive individual adopts the innovation with probability $p$ ($0<p<1$),
	and otherwise becomes informed. 
	
	\item An informed individual adopts the innovation upon encounter with an adopter.
	
	\item Demographic turnover (recruitment into and departure from the relevant population, for example through births and deaths) occurs at per capita rate $\mu$, 
	so that $\mu^{-1}$ is the mean residence time of an individual in the population ({\it{e.g.}}, the life expectancy). 
	Since recruitment and departure rates are assumed equal, we are assuming a constant population size.
	It is also assumed that individuals entering the population are naive (class $S_0$).
\end{itemize}

\begin{figure}
	\begin{center}
		\includegraphics[width=0.65\linewidth]{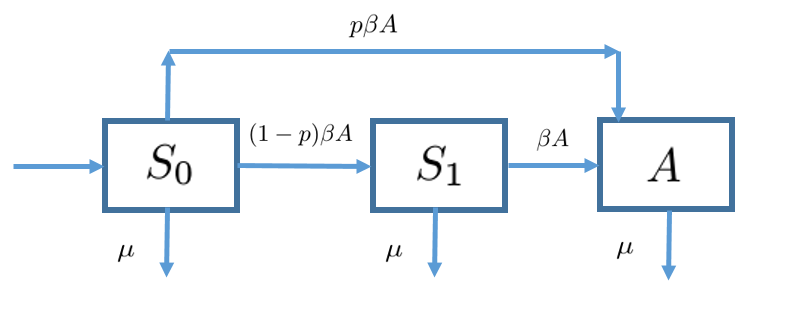}				
	\end{center}
	\caption{Diagram for the two-stage contagion model with permanent adoption}
	\label{d1}
\end{figure}

Denoting the fraction of the population in each of the three classes at time $t$ by $S_0(t)$, $S_1(t)$, $A(t)$ (so that
$S_0(t)+S_1(t)+A(t)=1$), the above assumptions, represented also in the diagram of figure \ref{d1}, 
translate, in the standard way \cite{keeling,martcheva}, into the following differential equations:
\begin{equation}\label{ss02}
S_0'=\mu-\beta S_0A-\mu S_0,
\end{equation}
\begin{equation}\label{ss12}
S_1'=(1-p)\beta S_0 A - \beta S_1 A-\mu S_1,
\end{equation}
\begin{equation}\label{aa2}
A'=\beta  [p S_0+ S_1]A - \mu A.
\end{equation}

\begin{figure}
	\begin{center}
		\includegraphics[width=0.85\linewidth]{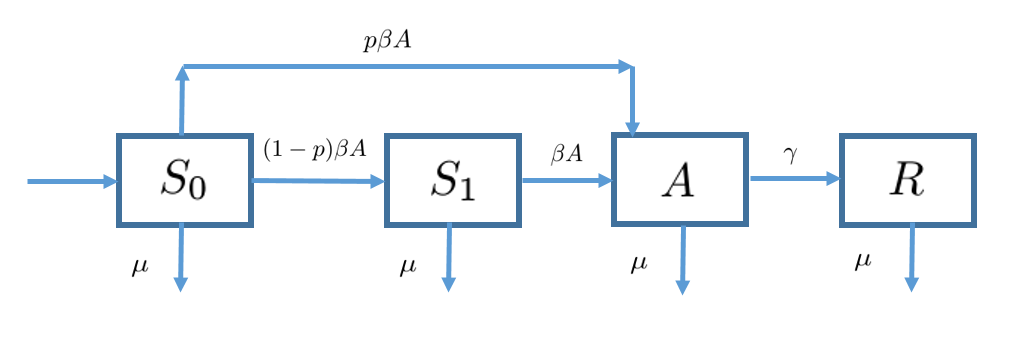}	
		
	\end{center}
	\caption{Diagram for the two-stage contagion model with temporary adoption}
	\label{d2}
\end{figure}

This is the {\it{permanent adoption}} model.
In our second model, the {\it{temporary adoption}} model, we assume that adopters abandon the innovation (or activists in a social movement become `burned out') at a constant per capita rate $\gamma$, so that the mean {\it{duration of adoption}} is $\gamma^{-1}$.
It is further assumed that those who have abandoned the innovation will not re-adopt it. This adds 
an additional class of removed individuals $R$ (see figure \ref{d2}), and changes the model equations to
\begin{equation}\label{s02}
S_0'=\mu-\beta S_0A-\mu S_0,
\end{equation}
\begin{equation}\label{s12}
S_1'=(1-p)\beta S_0 A - \beta S_1 A-\mu S_1,
\end{equation}
\begin{equation}\label{ax2}
A'=\beta  [p S_0+ S_1]A - (\gamma +\mu) A,
\end{equation}
\begin{equation}\label{r2}
R'=\gamma A -\mu R.
\end{equation}

It should be stressed here that a crucial feature of the two-stage contagion models is that both transitions between
stages depend on contagion. A model in which the transition from stage $S_1$ to stage $A$ were a spontaneous one, that is occuring 
at a constant per capita rate, would be equivalent to the standard SEIR model \cite{keeling,martcheva}, and would not generate any of the
interesting behaviors that are our focus here.

\subsection{Previous work on two-stage contagion models}
\label{previous}

We briefly survey previous work on two-stage contagion models of which we are aware, and compare the models 
which have been investigated with the model studied here.

In the statistical physics literature, a two-stage contagion model was introduced in \cite{janssen1,janssen2} under the
name Extended General Epidemic Process (EGEP). Like the model considered here, this model includes two stages, with contagion 
leading either directly to adoption ($A$) or to moving to a `weakened' stage ($S_1$), which upon further contagion can lead to adoption, and with permanent removal of 
adopters to a recovered stage (temporary adoption). An equivalent model is introduced in \cite{chen}, as a way to describe 
co-infection with two pathogens, under some symmetry assumptions.
A similar model was introduced in \cite{krapivski}, with the aim of describing innovations (with permanent adoption) and
fads (with temporary adoption), and including an arbitrary number of stages, though without direct movement of naive individuals 
into the adopter class. 
The models above do not include demographic turnover, so that there
is no mechanism for the renewal of the naive population, and they therefore generate transient epidemics rather than endemic states (except for
the permanent adoption model of \cite{krapivski}, in which the entire population eventually adopts).
Their investigation thus centers on the size of the epidemic, that is the total number of individuals who become infected 
before it fades.
Of prime interest here is the fact that, in contrast with the one-stage epidemic model, in the two-stage model one has, under certain conditions, a discontinuous (`first order' in the language of statistical physics) dependence of the total size of the epidemic on the contact parameter. This has led to interest in the statistical physics community, and several works investigate the two-stage contagion
process on lattices and random graphs \cite{bizhani,choi1,choi2,chung,hasegawa}.

In \cite{wang}, which develops models inspired by Rogers' Diffusion of Innovation Theory \cite{rogers}, mechanisms inducing renewal of the naive population are introduced, so that an endemic contagion becomes possible, as in this work. However, there are also 
some signficant differences between the assumptions in \cite{wang} and in our models, which we now detail.

In the first model of \cite{wang}, it is assumed that adoption is temporary, but the adopters who abandon the innovation return to the naive class so that they may re-adopt (this is analogous to the SIS model in epidemiology). In addition, informed individuals (class $S_1$) move back to the naive stage (`forgetting') at a constant per capita rate. These two flows are the mechanism providing renewal of the pool of naive individuals -- the model does not incorporate demographic turnover.
By contrast, in our model we do not allow individuals to move back into the naive stage, and the renewal of the naive pool 
is provided by demographic turnover. This is not a trivial difference, as may be seen from the fact that the model of \cite{wang}
reduces to a two-dimensional one, whereas our temporary adoption model is essentially three dimensional and cannot be reduced to 
a two-dimensional one. The appropriateness of one set of assumptions
or the other (or some different combination of assumptions) is dependent on the precise nature of the contagion phenomena involved, as 
well as on the relevant time scale.
One can construct a larger model including both models, but in order to understand the 
specific contribution of each mechanism there is much advantage in investigating simple models highlighting that specific mechanism.

The models of \cite{wang} also included the mechanisms of `spontaneous' transition of naive individuals to the informed 
stage and of informed individuals to adoption stage, at constant per capita rates, as a way to model the effects of mass-media.
Indeed for some of the results in \cite{wang} it was assumed that the mass-media effects are strong relative to 
the contagion effect. Here we treat the opposite extreme of pure contagion, without adding the media effects, as in
the previous works on the EGEP model. Since contagion effects are nonlinear while media effects are linear, some of the interesting 
nonlinear phenomena which follow from the contagion effects are  muted or cancelled when media effects are sufficiently strong. When media effects are present but sufficiently weak, behavior will be similar to the pure-contagion case, by structural stability. 
On the other hand, the model of \cite{wang} does not include direct contagion of naive individuals into the adopter
class. Here we do include this mechanism (controlled by the parameter $p$), as in the EGEP model, which is in fact
essential for the generation of some of the more interesting phenomena, such as the periodic oscillations.

A further element which is introduced in the first model of \cite{wang} is the possibility of a nonlinear dependence of the
per-capita transition rate from the informed class ($S_1$) to the adopter class on the number of adopters, that is replacing the term
$\beta A S$ by a term of the form $ g(A)S$, where $g$ is a nonlinear function. It is shown that choosing $g(A)$ to be quadratic leads to bistability.
In our models we show that bistability occurs even without this additional mechanism, and we will not introduce such nonlinear dependence.

The second model of \cite{wang} introduces a time delay to model the intermediate evaluation stage, as 
well as demographic turnover, though not the direct contagion of naive individuals into the adoption stage.
The time delay makes the stability analysis of equilibria considerably more difficult, but an interesting consequence of this 
delay is that it gives rise to periodic oscillations for some parameter values. Note that in our model with temporary adoption, 
we will show that periodic oscillations arise even in the absence of a delay. We will not consider delays in this work.

To conclude this short survey, we mention the work \cite{coletti}, which contains rigorous results on a stochastic two-stage 
contagion model on a lattice with return of adopters to the naive stage. In this work we restrict ourselves to mean-field deterministic  models.

\section{The permanent adoption model}
\label{permanent}

In this section we analyze the dynamics of the permanent adoption model. 
It will be useful to define the dimensionless {\it{contact parameter}} 
$$\delta=\frac{\beta}{\mu},$$
which plays a role similar to the basic reproductive number $R_0$ in epidemiological models: since an adopter spends 
an average time $\mu^{-1}$ in the adopter compartment before leaving the population, $\delta$ is the average total number of 
effective encounters that an adopter has.

We note that since $S_0+S_1+A=1$, we can eliminate one of the variables, say $S_1$, by substituting
\begin{equation}\label{ss1}S_1=1-S_0-A,\end{equation}
and reformulate the model equations (\ref{ss02})--(\ref{aa2}) as a two-dimensional system
\begin{equation}\label{es0}
S_0'=f(S_0,A)\doteq \mu-\beta S_0A-\mu S_0,
\end{equation}
\begin{equation}\label{ea}
A'=g(S_0,A)\doteq\beta [1-(1-p)S_0-A]A - \mu A.
\end{equation}
This defines a flow in the invariant region of the phase plane given by
$$S_0\geq 0,\;\; A\geq 0,\;\; S_0+A_0\leq 1.$$
We can use the Bendixon-Dulac criterion \cite{teschl} to verify that this system does not have limit cycles. Indeed
$$\frac{\partial}{\partial S_0}\left(\frac{1}{A}f(S_0,A) \right)+\frac{\partial}{\partial A}\left(\frac{1}{A}g(S_0,A) \right)
=-2\beta-\frac{\mu}{A}<0.$$
Since this is a two dimensional system in a bounded region, the Poincar\'e-Bendixon theorem \cite{teschl} implies that 
\begin{proposition}
	Every trajectory of the system (\ref{es0}),(\ref{ea}) approaches an equilibrium point as $t\rightarrow \infty$.
\end{proposition}
We therefore now study the equilibria of the model, which correspond to solutions of the algebraic equations
\begin{equation}\label{ees0}
\mu-\beta S_0A-\mu S_0=0,
\end{equation}
\begin{equation}\label{eea}
\beta [1-(1-p)S_0-A]A - \mu A=0.
\end{equation}

From (\ref{ees0}) we have
\begin{equation}\label{l1}S_0=\frac{1}{\delta A+1}.\end{equation}
From (\ref{eea}) we have
\begin{equation}\label{l2}A=0\;\;\;{\mbox{or}}\;\; (1-p)S_0+A = 1-\delta^{-1}.\end{equation}
If $A=0$ then we obtain the {\it{contagion-free}} equilibrium 
$$E_0: \;\; (S_0,S_1,A)=(1,0,0).$$
Assuming $A\neq 0$, and substituting (\ref{l1}) into (\ref{l2}) gives
$$\frac{1-p}{\delta A+1}+A = 1-\delta^{-1},$$
which may be rewritten as a quadratic equation and solved to give
\begin{equation}\label{exp}A_{1,2}=\frac{1}{2}\cdot\left[ 1-2\delta^{-1}\pm\sqrt{1-4(1-p)\delta^{-1}}\right].\end{equation}
These solutions will correspond to endemic equilibria if and only if they are real and positive.
We will denote by $E_1$ the equilibrium corresponding to $A=A_1$, with the other components given by (\ref{l1}),(\ref{ss1}), 
and by $E_2$ the equilibrium corresponding to $A_2$. We now determine the conditions on the parameters under which
these equilibria exist.

$A_1,A_2$ are real if and only if
\begin{equation}\label{cd0}\delta\geq 4(1-p).\end{equation}
Assuming now that (\ref{cd0}) holds, we have
\begin{equation}\label{ap10}A_1>0 \;\;\Leftrightarrow\;\;\sqrt{1-4(1-p)\delta^{-1}}> 2\delta^{-1}-1\;\;\Leftrightarrow\;\;\delta>2\;\; {\mbox{or}}\;\;\frac{1}{p}<\delta\leq 2.\end{equation}
Noting that $\frac{1}{p}\geq 4(1-p)$, we get 
$$A_1>0\;\;\Leftrightarrow\;\;\delta \geq \begin{cases}
4(1-p) & {\mbox{when}}\;\;p\leq \frac{1}{2}\\
\frac{1}{p} & {\mbox{when}}\;\;p> \frac{1}{2}
\end{cases}.$$
For the equilibrium $E_2$, we have, assuming (\ref{cd0}) holds,
\begin{equation}\label{am10}A_2> 0 \;\;\Leftrightarrow\;\;\sqrt{1-4(1-p)\delta^{-1}}< 1-2\delta^{-1}\;\;\Leftrightarrow\;\;2<\delta< \frac{1}{p},\end{equation}
so that 
$$A_2>0 \;\;\Leftrightarrow\;\;p<\frac{1}{2}\;\; {\mbox{and}}\;\; 4(1-p)<\delta< \frac{1}{p}.$$

To determine the stability of the equilibria, we linearize (\ref{es0}) around an equilibrium, obtaining the Jacobian matrix \cite{teschl}
$$J=\left(\begin{array}{cc}
-\beta A-\mu &  -\beta S_0\\ 
-\beta (1-p) A & \beta [1-(1-p)S_0]-2\beta A - \mu
\end{array}  \right)$$

For the contagion-free equilibrium $E_0$ we have
$$J=\left(\begin{array}{cc}
-\mu &  -\beta \\ 
0 & \beta p- \mu
\end{array}  \right),$$
with eigenvalues $-\mu,\beta p-\mu$, and we conclude that $E_0$ will be stable when both of these eigenvalues are
negative, that is when $\delta p<1$, and unstable if $\delta p>1$.

For the endemic equilibria $E_1,E_2$ we have, using (\ref{l2})
$$J=\left(\begin{array}{cc}
-\beta A-\mu &  -\frac{\beta}{1-p} (1-A-\delta^{-1})\\ 
-\beta (1-p) A & -\beta A 
\end{array}  \right)$$
so that
$$tr(J)= -2\beta A-\mu<0,\;\;\;\det(J)= \beta^2 A[ 2A+2\delta^{-1}-1] , $$
hence the equilibrium is stable if and only if $\det(J)>0$, that is iff
\begin{equation}\label{sstt} 2(A+\delta^{-1})> 1\end{equation}
Using the explicit solutions (\ref{exp}) we have that (\ref{sstt}) is equivalent to 
$$ 1\pm\sqrt{1-4(1-p)\delta^{-1}}>1,$$
which holds if and only if the $+$ sign is taken. 
We conclude that $E_1$ is stable whenever it exists, while $E_2$ is always unstable whenever it exists.

We summarize the results of the preceeding analysis.

\begin{proposition}\label{equi0}
	
	(I) If $p<\frac{1}{2}$ then:
	\begin{itemize}
		\item	For $\delta<4(1-p)$ there are no endemic equilibria, and the contagion-free equilibrium $E_0$ is 
		stable.
		
		\item For $4(1-p)< \delta<\frac{1}{p}$ there are two endemic equilibria $E_1,E_2$, with
		$E_1$ stable and $E_2$ unstable, and the contagion-free equilibrium $E_0$ is stable.
		
		\item For 	$\delta \geq \frac{1}{p}$ there is a unique endemic equilibrium $E_1$, which is stable, and the 
		contagion-free equilibrium $E_0$ is unstable.
	\end{itemize}
	
	(II) If $p\geq \frac{1}{2}$ then: 
	\begin{itemize}
		\item For $\delta\leq \frac{1}{p}$ there is no endemic equilibrium, and the contagion-free equilibrium $E_0$ is 
		stable.
		
		\item For $\delta>\frac{1}{p}$ there is a unique endemic equlibrium $E_1$, and the contagion-free equilibrium $E_0$ is 
		unstable.
	\end{itemize}
\end{proposition}

We now discuss the interesting dynamical consequences of the preceding results.

\begin{figure}
	\begin{center}
		\includegraphics[width=0.4\linewidth]{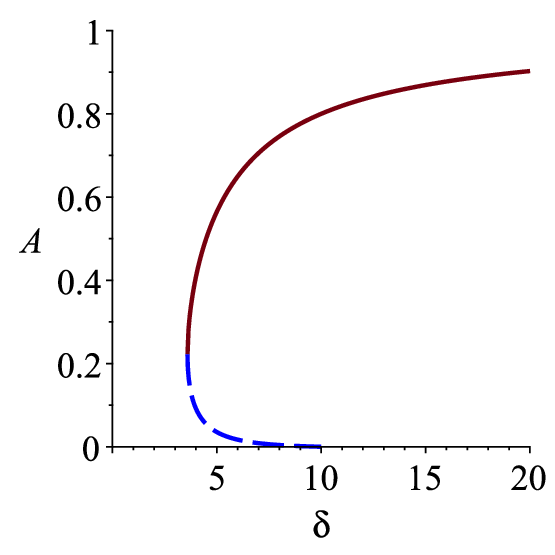}	
		\includegraphics[width=0.4\linewidth]{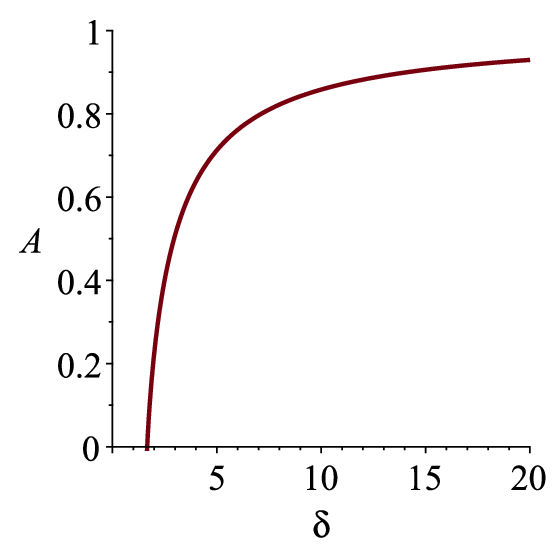}	
	\end{center}
	\caption{Fraction of adopters $A$ at endemic equilibria, as a function of $\delta$, for $p=0.1$ (left) and for $p=0.6$ (right).
		The red (full) is the equilibrium $E_1$, which is stable, and the blue (dashed) line in the left part is the
		unstable equilibrium $E_2$.}
	\label{bb0}
\end{figure}

When $p\geq \frac{1}{2}$ (see figure \ref{bb0}, right), the behavior is similar to that familiar from a standard one-stage 
contagion model -- the SI
model with demographic turnover: if contact rate is low ($\delta<\frac{1}{p}$) then contagion cannot spread 
(the contagion-free equilibrium is stable, and there is no endemic equilibrium), and as $\delta$ crosses the {\it{invasion threshold}}
$\delta=\frac{1}{p}$ the contagion-free equilbrium loses stability and an endemic equilibrium is born, so that
contagion is established. This transition is a {\it{continuous}} one: for values of $\delta$ slightly above the
threshold, the fraction of adopters $A_1$ is small, and it increases as $\delta$ increases.

\begin{figure}
	\begin{center}
		\includegraphics[width=0.5\linewidth]{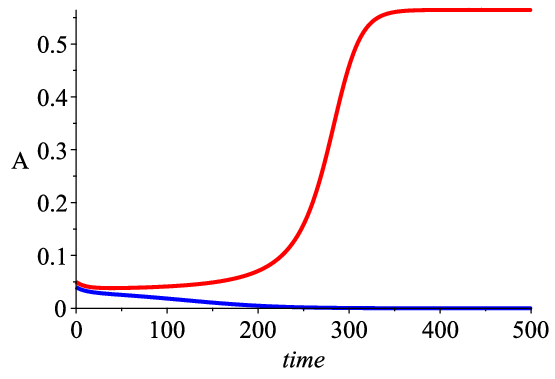}	
	\end{center}
	\caption{Fraction of adopters $A(t)$, for $p=0.1$, $\mu=0.05$, $\beta=0.25$ (so $\delta=5$), for initial conditions:
		$A(0)=A_0,S_0(0)=1-A_0,S_1(0)=0$, $A_0=0.04,A_0=0.05$.}
	\label{bb}
\end{figure}

Things are more interesting in the case $p<\frac{1}{2}$ (see figure \ref{bb0}, left), since in this case we have a {\it{critical transition}} \cite{scheffer} 
at the {\it{endemicity threshold}} $\delta=4(1-p)$, in which two endemic equilibria $E_1,E_2$ appear (`out of the blue'), so that contagion can establish at the level
$A_1$ corresponding to the stable equilibrium $E_1$. At the endemicity threshold $\delta=4(1-p)$ we have
\begin{equation}\label{ja}A_1= \frac{1}{2}\cdot\left[ 1-\frac{1}{2(1-p)}\right]>0\end{equation}
so that as this threshold is crossed the level of contagion can jump from $0$ to a positive value.
Note however that even above the endemicity threshold, the contagion-free 
equilibrium remains stable until the invasion threshold $\delta=\frac{1}{p}$ (which is larger) is crossed. This means
that for values of $\delta$ between these two thresholds we have {\it{bistability}} - contagion may establish, or not, depending on whether the initial conditions belong to the basin of attraction of $E_0$ or of $E_1$. This is demonstrated in figure \ref{bb}, in which we show the solution $A(t)$ for 
parameter values $p=0.1,\delta=5$, for two initial conditions: when the initial fraction of adopters is $4\%$ the 
contagion is extinguished, while for an initial fraction $5\%$ contagion is established, with equilibrium value of $A_1=56.5\%$ of the
population. Thus under essentially the same conditions - that is the same parameter values and only slightly different
initial conditions, the system may achieve radically different outcomes. 

The critical transition displayed by this model has important implications with regard to 
changes in outcomes under variation of its parameters. 
If we assume that initially we are near the contagion-free equilibrium $E_0$, with $\delta<4(1-p)$, and slowly
increase the contact parameter $\delta$ ({\it{e.g.}} by increasing the contact rate $\beta$)
then, as the endemicity threshold $\delta=4(1-p)$ is crossed we will still be in the basin of attraction of the $E_0$, 
so that the population will remain contagion-free. This will continue until the invasion threshold $\delta=\frac{1}{p}$ is reached,
at which point $E_0$ loses stability, and then we will observe an even more dramatic jump to the value 
$A_1$ corresponding to $\delta =\frac{1}{p}$, that is to 
$$A_1=\frac{1}{2}\cdot\left[ 1-2p+\sqrt{1-4(1-p)p}\right].$$
On the other hand, if initially $\delta>\frac{1}{p}$ and we are near the endemic equilibrium $E_+$ and slowly decrease
$\delta$, then we will remain near the endemic equilibrium even as $\delta$ drops below the invasion threshold and the 
contagion-free equilibrium becomes stable. This will continue until the endemicity threshold $\delta=4(1-p)$ is 
reached, at which point the endemic equilibria disappear and we will jump from the value given by (\ref{ja}) to a contagion-free state.
We thus have the phenomenon of {\it{hysteresis}} in which discontinuous transitions occur at different values 
of the parameter, depending on whether it is increased or decreased. 

In the extreme case $p=0$ (no direct adoption by naive individuals), there is no invasion threshold and the 
contagion-free equilibrium is stable for all $\delta$. In this two endemic equilibria are born when $\delta=4$, and we have bistability for all $\delta>4$.

\begin{figure}
	\begin{center}
		\includegraphics[width=0.4\linewidth]{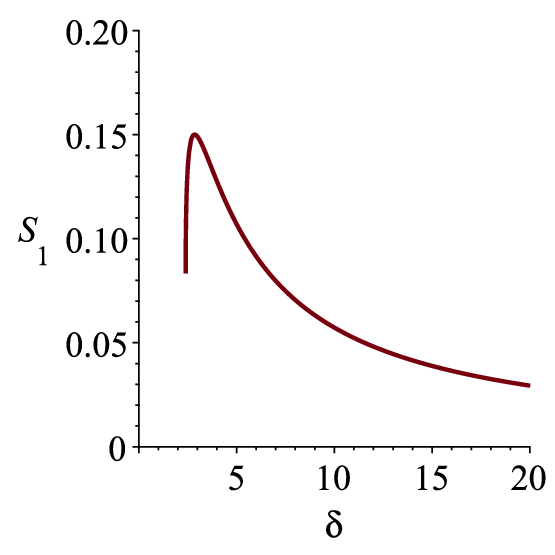}	
		\includegraphics[width=0.4\linewidth]{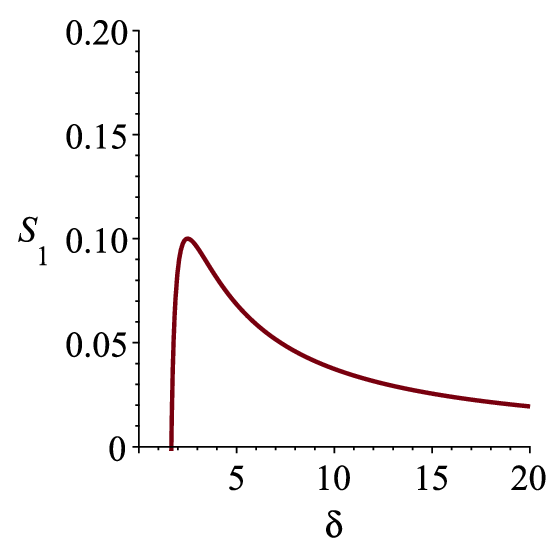}	
	\end{center}
	\caption{Fraction of informed individuals $S_1$ at the stable equilbirium $E_1$, as a function of $\delta$, for $p=0.4$ (left) and for $p=0.6$ (right).}
	\label{s1plot}
\end{figure}

To conclude our discussion of the permanent adoption model, it is of interest to consider the dependence of the 
sizes of the different population fractions at the stable equilibrium $E_1$ on the contact parameter $\delta$.
$A=A_1$ is given by (\ref{exp}), and by (\ref{ss1}),(\ref{l1}), we have
$$S_0= \frac{1-\sqrt{1-4(1-p)\delta^{-1}}}{ 2(1-p)},\;\;\;S_1=\delta^{-1}- \frac{p(1-\sqrt{1-4(1-p)\delta^{-1}})}{ 2(1-p)}.$$ 
While it is immediate that $S_0$ is monotone decreasing and $A$ is monotone increasing as a function $\delta$,
the fraction $S_1$ of individuals at the intermediate stage is not monotone in $\delta$, as shown in figure 
\ref{s1plot}. Indeed it is easy to calculate that the fraction $S_1$ is maximized at $\delta=\frac{4}{p+1}$, attaining the value
$S_1=\frac{1-p}{4}$.

\section{The temporary adoption model}
\label{temporary}

We now move to the analysis of the two-stage contagion model with temporary adoption, which displays a richer 
repertoire of dynamic behaviors.
As for the previous model, it will be useful to define the non-dimensional parameter
$$\delta=\frac{\beta}{\gamma+\mu},$$
which can again be interpreted as the mean number of effective contacts that an adopter makes during the period of adoption, so that 
it will be called the contact parameter.

\subsection{Analysis of equilibria}
\label{equilibria}

\subsubsection{Existence of equilibria}

Equilibria of the model are given by solutions, with non-negative components, of the equations obtained by equating the
derivatives in (\ref{s02})-(\ref{r2}) to zero:
\begin{equation}\label{e1}\mu-\beta S_0A-\mu S_0=0,\end{equation}
\begin{equation}\label{e2}(1-p)\beta S_0 A - \beta S_1 A-\mu S_1=0,\end{equation}
\begin{equation}\label{e3}\beta A [p S_0+ S_1] - (\gamma +\mu) A=0,\end{equation}
\begin{equation}\label{e4}\gamma A -\mu R=0.\end{equation}
We now analyze the solutions of this algebraic system.

From (\ref{e1}),(\ref{e2}),(\ref{e4}) we have
\begin{equation}\label{it}S_0= \frac{1}{\delta (\frac{\gamma}{\mu}+1) A+ 1},\;\;S_1=(1-p)\cdot \frac{\delta(\frac{\gamma}{\mu}+1)  A}{(\delta(\frac{\gamma}{\mu}+1) A+ 1)^2},\;\;
R=\frac{\gamma}{\mu}\cdot A.
\end{equation}

From (\ref{e3}) we have that either $A=0$ or
\begin{equation}\label{an0} p S_0+ S_1 = \delta^{-1}  .\end{equation}
In the case $A=0$ we obtain the contagion-free equilibrium 
\begin{equation}\label{cf} E_0:\;\;\;A=0,\;\;S_0=1,\;\;S_1=0,\;\;R=0.\end{equation}
In the case $A\neq 0$ we have, substituting (\ref{it}) into (\ref{an0}),
$$p \cdot \frac{1}{(\frac{\gamma}{\mu}+1) A+ \delta^{-1}}+(1-p)\cdot \frac{(\frac{\gamma}{\mu}+1)  A}{((\frac{\gamma}{\mu}+1) A+ \delta^{-1})^2} =1, $$
which is equivalent to a quadratic equation with solutions
\begin{equation}\label{exp1}A_{1,2}=\frac{1}{2}\cdot \left(1+\frac{\gamma}{\mu}\right)^{-1}\left[ 1-2\delta^{-1}\pm\sqrt{1-4(1-p)\delta^{-1}}\right].
\end{equation}
The expression (\ref{exp1}) is identical to the expression (\ref{exp}) for the permanent 
adoption model, apart from the multiplicative factor $\left(1+\frac{\gamma}{\mu}\right)^{-1}$, (indeed when 
$\gamma=0$ the temporary adoption model degenerates to the permanent adoption model). Therefore the analysis of the conditions
for existence of equilibria is the same in both cases, and we obtain:

\begin{proposition}\label{equi}
	
	(I) If $p<\frac{1}{2}$ then:
	\begin{itemize}
		\item	For $\delta<4(1-p)$ there are no endemic equilibria, 
		
		\item For $4(1-p)\leq \delta<\frac{1}{p}$ there are two endemic equilibria $E_1,E_2$ (which coincide when $\delta=4(1-p)$).
		
		\item For 	$\delta \geq \frac{1}{p}$ there is a unique endemic equilibrium $E_1$.
	\end{itemize}
	
	(II) If $p\geq \frac{1}{2}$ then: 
	\begin{itemize}
		\item For $\delta\leq \frac{1}{p}$ there is no endemic equilibrium,
		
		\item For $\delta>\frac{1}{p}$ there is a unique endemic equlibrium $E_1$.
	\end{itemize}
\end{proposition}

However, the conditions for stability of the equilibria, which to a large extent determine the dynamics of the model,
are quite different from those for the permanent adoption model, and we turn to these next.

\subsubsection{Stability of equilibria}

To investigate stability of the equilibria, we examine the linearization of the system around an equilibrium \cite{teschl}.
In fact since $R$ does not appear in the first three equations, and is determined by the other variables by
$R=1-S_0-S_1-A$, it suffices to consider (\ref{s02})-(\ref{ax2}). Linearization of this system around an 
equilibrium $(S_0,S_1,A)$ gives the Jacobian matrix
$$J=\left(\begin{array}{ccc}
-\beta A -\mu & 0 & -\beta S_0 \\ 
(1-p)\beta A &  -\beta  A-\mu  & (1-p)\beta S_0  - \beta S_1  \\ 
p \beta A   & \beta A  & \beta  [p S_0+ S_1] - \gamma -\mu 
\end{array}  \right).$$
Beginning with the contagion-free equilibrium (\ref{cf}), we have
$$J=\left(\begin{array}{ccc}
-\mu & 0 & -\beta  \\ 
0 & -\mu    & (1-p)\beta    \\ 
0  & 0 & \beta  p  - \gamma -\mu 
\end{array}  \right)$$
With eigenvalues $\lambda_1=\lambda_2=-\mu$ and $\lambda_3=\beta  p  - \gamma -\mu $, so that 
we obtain
\begin{proposition}\label{cfs}
	The contagion-free equilibrium $E_0$ is stable if and only if 
	$\delta<\frac{1}{p}$.
\end{proposition}

Moving to the endemic equilibria, we compute the characteristic polynomial of $J$, and substitute the expressions 
(\ref{it}) for $S_0,S_1$ at the equilibrium, to obtain
\begin{equation}\label{cp}P(\lambda)=\lambda^3+a_2\lambda^2+a_1\lambda+a_0,\end{equation}
where
\begin{equation}\label{a0}a_0=\beta\mu\left( (1-\delta^{-1})(\beta A+\mu)+(1-p )\mu\left(\frac{\mu}{\beta A+ \mu} -  2\right)\right)\end{equation}
\begin{equation}\label{a1}a_1=   ( \beta  A+\mu)^2+(p-\delta^{-1})\beta\mu\end{equation}
\begin{equation}\label{a2}a_2= 2(\beta  A +\mu ).\end{equation}
By the Routh-Hurwitz stability criterion \cite{teschl}, the equilibrium is stable when:
$$a_0>0,\;\;a_2>0,\;\;a_1a_2>a_0.$$

The condition $a_2>0$ holds automatically. To check the condition $a_0>0$, we substitute (\ref{exp1}) into (\ref{a0}) and obtain
(with a $+$ sign for $A_1$ and a $-$ sign for $A_2$):
\begin{equation}\label{a01}a_0=\frac{1}{2}\cdot \beta\mu^2 \delta \left(\sqrt{1-4(1-p)\delta^{-1}} \pm (1-2\delta^{-1}) \right)\sqrt{1-4(1-p)\delta^{-1}},\end{equation}
whence
$$a_0>0\;\;\Leftrightarrow\;\;\sqrt{1-4(1-p)\delta^{-1}} \pm (1-2\delta^{-1})>0.$$
For $A=A_1$ ($+$ sign), the last condition always holds, while for $A=A_2$ it is equivalent to 
$$\sqrt{1-4(1-p)\delta^{-1}} > 1-2\delta^{-1}\;\;\Leftrightarrow\;\;  \delta > \frac{1}{p},$$
but under the last condition $A_1$ is the unique endemic equilibrium, so we conclude that the condition $a_0>0$ never holds for 
$A=A_2$. 

We have therefore shown that
\begin{proposition}\label{e2s}
	The equilibrium $E_2$, when it exists, is unstable.
\end{proposition}

By the above we have that $A=A_1$ will be stable if and only if $a_1a_2>a_0$, and we proceed to check when this condition holds.
Substituting $A=A_1$ as given by (\ref{exp1}) into (\ref{a0})-(\ref{a2}) we calculate
$$a_1a_2-a_0= \frac{1}{4}\cdot\mu^3 \delta^3 \left[ 1+ \sqrt{1-4(1-p)\delta^{-1}}\right]^3+\beta\mu^2\delta(p-\delta^{-1})\left[ 1+ \sqrt{1-4(1-p)\delta^{-1}}\right]$$
$$-\frac{1}{2}\cdot \beta\mu^2 \delta \left(\sqrt{1-4(1-p)\delta^{-1}} + (1-2\delta^{-1}) \right)\sqrt{1-4(1-p)\delta^{-1}},$$
so that $a_1a_2>a_0$ is equivalent to 
\begin{equation}\label{sh}\frac{\mu}{\gamma+\mu} \delta^2 \left[ 1+ \sqrt{1-4(1-p)\delta^{-1}}\right]^3>2(1-2p)\cdot  \left((\delta-2) + \delta\sqrt{1-4(1-p)\delta^{-1}} \right).\end{equation}
We note that if $p\geq \frac{1}{2}$ then (\ref{sh}) automatically holds, since the left hand side is positive and the right hand side
is non-positive. If $p<\frac{1}{2}$ then (\ref{sh}) is equivalent to:
\begin{equation}\label{stb}\frac{\gamma}{\mu}<F(\delta)\doteq \frac{\delta^2}{2(1-2p)}\cdot  \frac{\left[ 1+ \sqrt{1-4(1-p)\delta^{-1}}\right]^{3}}{(\delta-2) + \delta\sqrt{1-4(1-p)\delta^{-1}} }-1.\end{equation}
We therefore conclude that
\begin{proposition}\label{e1s}
	(i) If $p\geq \frac{1}{2}$ then the endemic equilibrium $E_1$ is stable.
	
	(ii) If $p<\frac{1}{2}$ then the endemic equilibrium $E_1$ is stable if $\frac{\gamma}{\mu}<F(\delta)$ and 
	unstable if $\frac{\gamma}{\mu}>F(\delta)$, where $F(\delta)$ is defined in (\ref{stb}).
\end{proposition}

Since the function $F(\delta)$ plays an important role, we wish to understand the shape of its graph, assuming $p<\frac{1}{2}$. 
This function is defined for $\delta>4(1-p)$, and we have 
$$F(4(1-p))=\left(\frac{1}{1-2p}+1\right)^2-1,\;\;\;\lim_{\delta\rightarrow \infty}F(\delta)=+\infty.$$
Differentiating $F$, we find that
$$F'(\delta)>0\;\;\Leftrightarrow\;\; \frac{1}{2}\left(\delta-2\right)\sqrt{1-4(1-p)\delta^{-1}}
>2-p-\frac{1-p}{\delta}-\frac{\delta}{2},$$
which, using some elementary algebra, gives 
\begin{proposition}\label{Fp}
	(i) If $p<\frac{1}{4}$ then $F(\delta)$ is monotone increasing for all $\delta\geq 4(1-p)$.
	
	(ii) If $\frac{1}{4}\leq p<\frac{1}{2}$ then $F(\delta)$	is decreasing for $4(1-p)\leq \delta < 1+\frac{1}{\sqrt{p}}$ and increasing for 
	$\delta>1+\frac{1}{\sqrt{p}}$.
\end{proposition}

\subsection{The phase diagram: dynamics of the model}

We now synthesize our previous result, to obtain a picture of the dependence of the model dynamics on the 
parameters.

We first note that when $p\geq \frac{1}{2}$, the behavior is rather simple: for $\delta<\frac{1}{p}$ we have 
only the contagion-free equilibrium $E_0$, which is stable. At $\delta=\frac{1}{p}$ the contagion-free equilibrium 
loses stability and a stable endemic equilibrium $E_1$ arises in a continuous transition. This behavior 
is similar to that observed in the standard SIR model with demographic turnover \cite{keeling,martcheva}.

\begin{figure}
	\begin{center}
		\includegraphics[width=0.45\linewidth]{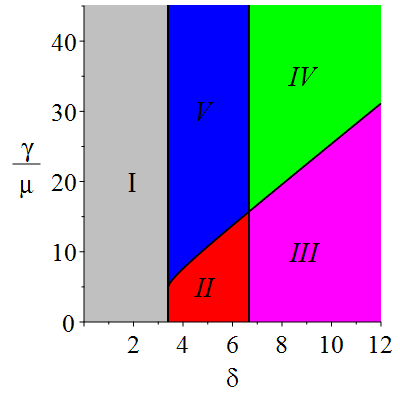}
		\includegraphics[width=0.45\linewidth]{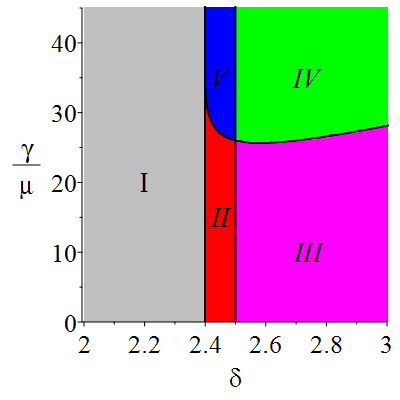}
		
	\end{center}
	\caption{Phase diagram for $p=0.15$ (left) and for $p=0.4$ (right).}
	\label{ppp}
\end{figure}

In the case $p<\frac{1}{2}$ we observe new phenomena.
To understand these, we divide the plane of parameters $\left(\delta,\frac{\gamma}{\mu}\right)$
into five regions, each of which corresponds to different properties of the equilibria, as determined in section
\ref{equilibria}.
These regions are shown in figure \ref{ppp} for the case $p=0.15$ and  $p=0.4$ (the interval between the endemicity threshold and
the invasion threshold is much narrower in the latter case, so we choose a different scale on the $\delta$-axis). The qualitative difference in appearance 
between the two cases stems from the fact that the function $F(\delta)$, which defines the boundary $\frac{\gamma}{\mu}=F(\delta)$
between regions $II$ and $V$ and between region $III$ and $IV$ is monotone increasing for when $p<\frac{1}{4}$ and has a minimum
when $p>\frac{1}{4}$ (proposition \ref{Fp}).

We now study the dynamics for each of these regions in turn, using both the analytical results regarding the 
equilibria and their stability obtained above and numerical simulations.

\subsubsection{Region $I$: No contagion}
In region $I$, defined by the condition
$$\delta <4(1-p),$$
the contagion-free equilibrium $E_0$ is stable, and there exist no endemic equilibria (propositions \ref{equi},\ref{cfs}). When the parameters are in this region 
the contagion will not spread.

\subsubsection{Region $II$: Bistability}
\begin{figure}
	\begin{center}
		\includegraphics[width=0.45\linewidth]{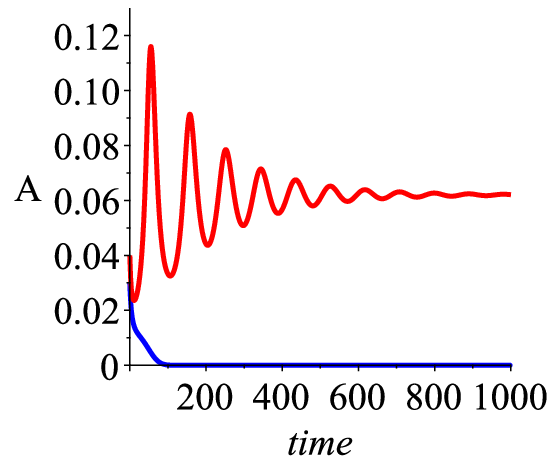}
		\includegraphics[width=0.45\linewidth]{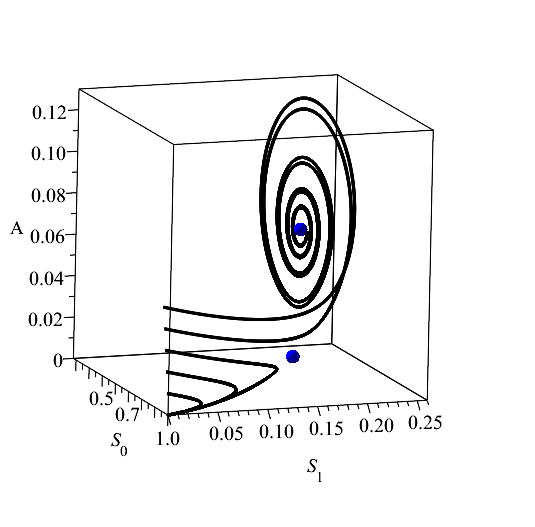}		
	\end{center}
	\caption{Dynamics for parameter values in region $II$. Left: Solution $A(t)$ for parameters $p=0.15,\mu=0.05,\delta=3.7,\frac{\gamma}{\mu}=5$. Initial conditions
		are $A(0)=A_0,\;\;S(0)=1-A_0,\;S_1=R=0$, with $A_0=0.04$, $A_0=0.03$. Right: Some trajectories in the phase space, for the 
		same parameter values, and initial conditions $A(0)=A_0,\;\;S(0)=1-A_0,\;S_1=R=0$ with $A_0=0.01,0.02,0.03,0.04,0.05$.
		This demonstrates the bistability of contagion-free and the endemic equilibrium $E_1$, when parameters are in region II. The endemic
		equilibria are shown in blue.}
	\label{II}
\end{figure}

In region $II$, given by the conditions
$$4(1-p)<\delta< \frac{1}{p},\;\; \frac{\gamma}{\mu}<F(\delta),$$
the contagion-free equilibrium $E_0$ is stable, but there exist also two endemic equilibria $E_1,E_2$, with 
$E_1$ stable and $E_2$ unstable (propositions \ref{equi},\ref{e2s},\ref{e1s}). Thus in this region we have {\it{bistability}} - the contagion may persist or not, 
depending on the initial conditions. This is demonstrated in figure \ref{II} (left), in which, when the initial fraction of adopters is $3\%$ the contagion dies out, while if the initial fraction of adopters is $4\%$ the contagion persists and approaches an endemic equilibrium.

Let us note that in this region the coefficients of the characteristic polynomial (\ref{cp}) of the linearization at $E_2$ satisfy
$a_0<0$, $a_2>0$. $a_0<0$ implies that the product of eigenvalues is positive, so 
there are either three positive eigenvalues, one positive eigenvalue and two negative eigenvalues, or one positive eigenvalue and
two complex conjugate eigenvalues. But
$a_2>0$ implies that the sum of all eigenvalues is negative, ruling out the possibility of three positive eigenvalues, and 
implying that if there are two complex conjugate eigenvalues then their real part must be negative. Therefore we conclude that $E_2$
is a saddle with a one-dimensional unstable manifold.

\subsubsection{Region $III$: Endemic equilibrium or bistability of equilbrium and limit cycle}
\begin{figure}
	\begin{center}
		\includegraphics[width=0.45\linewidth]{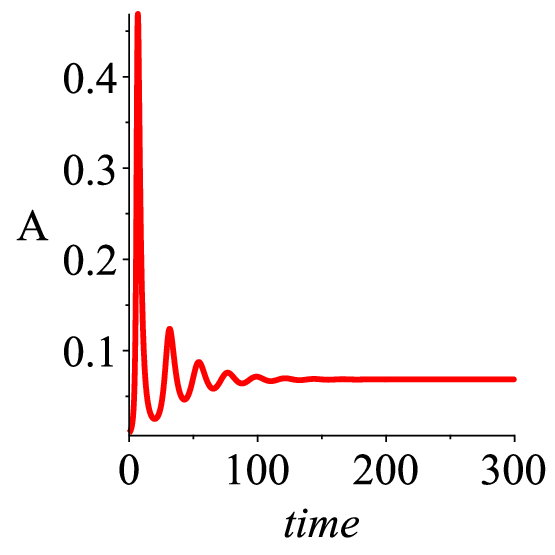}
		\includegraphics[width=0.45\linewidth]{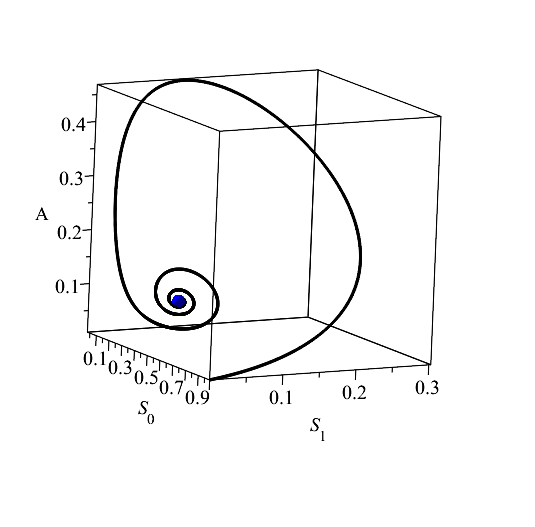}
		
	\end{center}
	\caption{Dynamics for (most) parameter values in region $III$. Left: Solution $A(t)$ for parameters $p=0.15,\mu=0.05,\delta=8,\frac{\gamma}{\mu}=10$. Right: A trajectory in the phase space, for the 
		same parameter values.}
	\label{III}
\end{figure}

In region $III$, given by the conditions
$$\delta>\frac{1}{p},\;\;\frac{\gamma}{\mu}<F(\delta),$$
the contagion-free equilibrium $E_0$ is unstable, and there exists a unique endemic equilibrium $E_1$, which is stable 
(propositions \ref{equi},\ref{e1s}).
Thus in this region contagion will not fade out, and it can persist at the stable endemic equilibrium. The simulations show
that this is indeed the case for most parameter values in region $III$, as illustrated in figure \ref{III}. 
However we also find a small range of parameter values in region $III$, near its boundary with region $IV$, for which 
a stable limit cycle coexists with the stable equilibrium $E_1$. For these parameter values we have bistabilty of 
the endemic equilibrium and a limit cycle, so that contagion can persist either at a constant or at periodically varying 
prevalence, depending on initial conditions. This phenomenon will be explained in section \ref{hopf}, when we discuss the 
Hopf bifurcation at the boundary of regions $III$ and $IV$.

\subsubsection{Region $IV$: Endogenous oscillations}

In region $IV$, given by the conditions
$$\delta>\frac{1}{p},\;\;\frac{\gamma}{\mu}>F(\delta),$$
the contagion-free equilibrium $E_0$ is unstable, and there exists a unique endemic equilibrim $E_1$, but it too
is unstable (propositions \ref{equi},\ref{e1s}). Indeed in this region we have that the coefficients of the characteristic polynomial (\ref{cp}) of the linearization at 
$E_1$ satisfy: $a_0>0,a_2>0,a_1a_2<a_0$. The fact that $a_1a_2<a_0$ implies that at least one eigenvalue of the linearization has
a positive real part. $a_0>0$ implies that the product of eigenvalues is negative, so that there is exactly one real negative eigenvalue,
and the other eigenvalues are either both positive or complex conjugate with a postive real part. In both cases we conclude that
the unstable manifold of $E_1$ is two-dimensional.

\begin{figure}
	\begin{center}
		\includegraphics[width=0.45\linewidth]{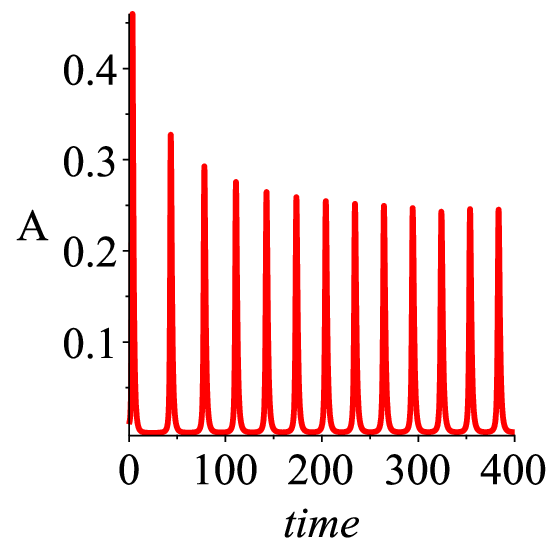}
		\includegraphics[width=0.45\linewidth]{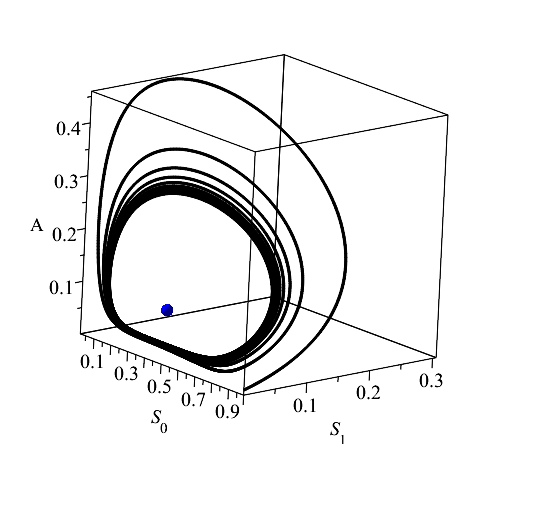}		
	\end{center}
	\caption{Dynamics for parameter values in region IV. Left: Solution $A(t)$ for parameters $p=0.15,\mu=0.05,\delta=8,\frac{\gamma}{\mu}=20$. Right: A trajectory in the phase space, for the 
		same parameter values.}
	\label{IV}
\end{figure}

\begin{figure}
	\begin{center}
		\includegraphics[width=0.32\linewidth]{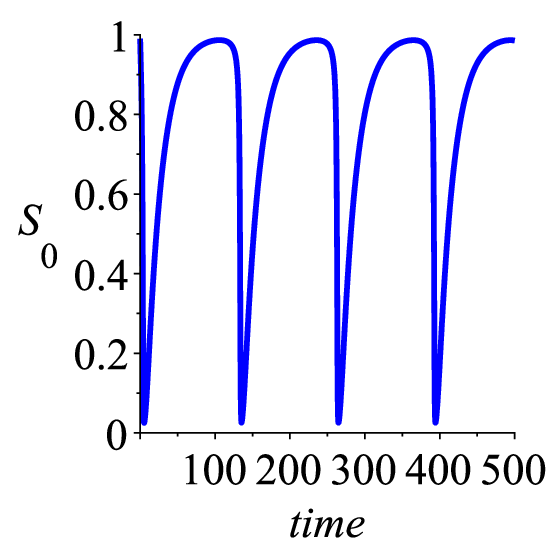}
		\includegraphics[width=0.32\linewidth]{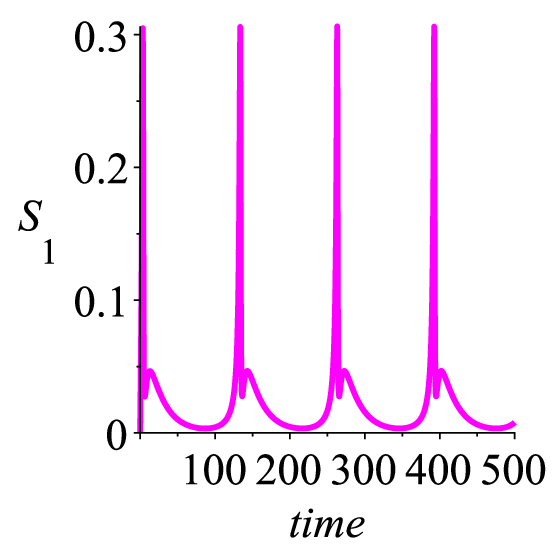}
		\includegraphics[width=0.32\linewidth]{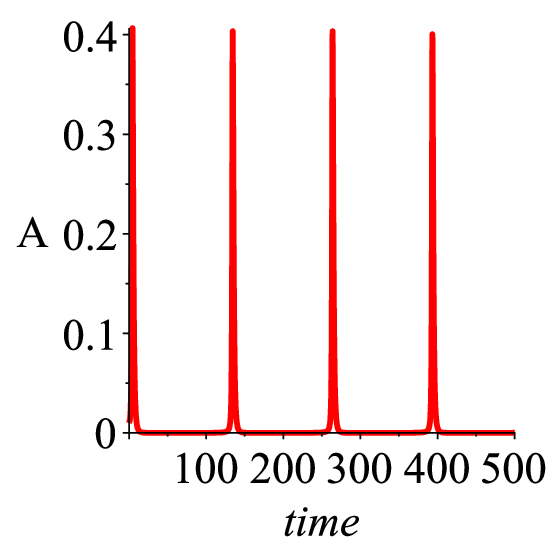}\\
	\end{center}
	\caption{Solution $A(t),S_0(t),S_1(t)$ for parameters $p=0.15,\mu=0.05,\delta=7,\frac{\gamma}{\mu}=20$.}
	\label{IV2}
\end{figure}

In view of the fact that the contagion-free equilibrium is unstable, we know that the contagion cannot fade out. On the other hand,
since  both equilibria $E_0,E_1$ are unstable, the system cannot stabilize at an equilibrium, and we conclude that the 
contagion must persist in a non-stationary regime. Simulations show that contagion persists as a
limit cycle, leading to sustained oscillations, representing repeated epidemic cycles.
This is 
demonstrated in figure \ref{IV}.
In this example $\delta=8$, the population turnover time is $\mu^{-1}=20$ years and the mean duration of adoption is $\gamma^{-1}=1$ year, and 
periodic oscillations have a period of around $29$ years are obtained, with the fraction of adopters varying
between less than $0.2\%$ to nearly $25\%$ of the population. The period of oscillations varies widely with the parameters.
In figure \ref{IV2} we display the oscillations where we have reduced the contact parameter to $\delta=7$ (by reducing $\beta$), keeping the same population turnover time and mean duration of adoption. The period of oscillations is now approximately $128$ years.

It should be noted that although the stability analysis of the equilibrium points showed that contagion must persist in a 
non-stationary state, it does not follow automatically that this must be a periodic one - indeed it is known that 
a three dimensional system can also exhibit quasi-periodic and chaotic behaviors. However, our simulations for various 
parameter values in region $IV$ have not revealed 
any non-stationary dynamics other than a limit cycle. Verifying this mathematically appears to be a challenging problem.

\subsubsection{Region V: excitability}
\label{excite}

\begin{figure}
	\begin{center}
		\includegraphics[width=0.45\linewidth]{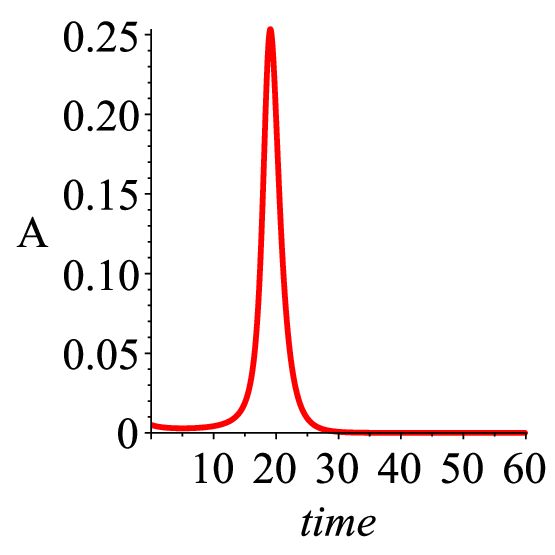}
		\includegraphics[width=0.45\linewidth]{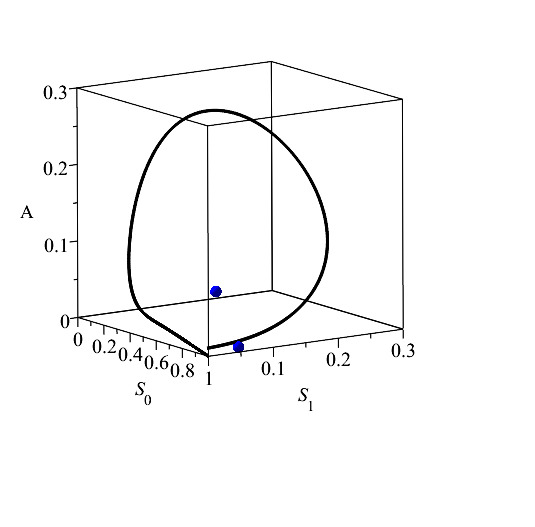}
	\end{center}
		\caption{Dynamics for parameter values in region V. Left: Solution $A(t)$ for parameters $p=0.15,\mu=0.05,\delta=5,\frac{\gamma}{\mu}=20$. Right: A trajectory in the phase space, for the 
			same parameter values.}
	\label{V}
\end{figure}

\begin{figure}
	\begin{center}
		\includegraphics[width=0.6\linewidth]{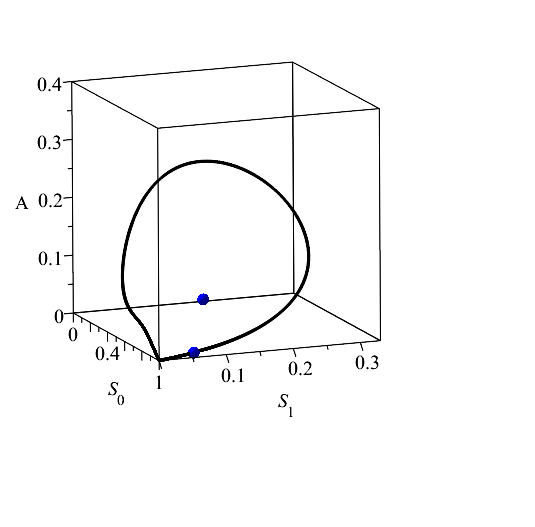}
	\end{center}
		\caption{Heteroclinic 
			loop involving the stable contagion-free equilibrium $E_0$ and the unstable endemic equilibrium $E_2$,
			for parameters $p=0.15,\mu=0.05,\delta=5,\frac{\gamma}{\mu}=20$ (equlibrium $E_1$ is also plotted, but not involved). }
	
	\label{het}
\end{figure}

In region $V$, given by the conditions
$$4(1-p)<\delta<\frac{1}{p},\;\;\frac{\gamma}{\mu}>F(\delta),$$
the contagion-free equilibrium $E_0$ is stable, and there exist two endemic equilibria $E_1,E_2$, both of 
which are {\it{unstable}} (propositions \ref{equi},\ref{e2s},\ref{e1s}). Thus in this region the contagion cannot persist in the form of an endemic equilibrium.
A priori one could think that contagion might persist in the form of endogenous oscillations (as is the case for region IV),
but the simulations show that this is not the case, and in fact generic trajectories converge to the contagion-free 
equilibrium as $t\rightarrow \infty$. However, here we observe a different phenomenon (see figure \ref{V}): trajectories in phase space, starting at points 
near the contagion-free equilibrium, make a large excursion away from this equilibrium and then back to it. This means that contagion 
spreads as a large epidemic, and then disappears. This is quite different from the behavior in region $I$, where contagion
disappears without spreading. This phenomenon is known as {\it{excitability}}, and it is familiar in the field of neuroscience 
\cite{izhikevich}. To understand the underlying reason for it, we need to look at the unstable manifold of the 
equilibrium $E_2$. Note that, as is the case for region $II$, since the coefficients of the characteristic polynomial of the linearization at $E_2$ satisfy $a_0<0,a_2>0$ we have that this unstable manifold is one-dimensional. When we plot this unstable manifold (by numerically solving the system, taking initial values very close to $E_2$), see figure \ref{het}, we observe that its two ends connect $E_2$ to the stable contagion-free equilibrium $E_0$, forming a heteroclinic cycle. This cycle attracts nearby trajectories and
is responsible for the excitability phenonmenon. As we will see further on, the heteroclinic cycle above can be understood as 
a `residue' of the limit cycle which exists when parameters are in region $IV$, arising from it through a homoclinic bifurcation.

\subsection{Bifurcations}

Having characterized the model dynamics in each of the five regions of the parameter plane, it is interesting 
to understand the transitions that occur when the boundaries from one region to another are crossed. Each such crossing 
corresponds to a bifurcation involving one of the equilibrium points.
We will examine the six transitions that can occur
when the parameters vary along a generic curve in the parameter plane. 
To illustrate this, we take $p=0.15$ (the corresponding phase diagram is in figure \ref{ppp}, left), choose three horizontal lines in the $(\delta,\frac{\gamma}{\mu})$-plane,
($\frac{\gamma}{\mu}=4,10,20$) and examine the bifurcations along these lines, as they cross regions. The $A$-values of the equilibria, as well
the range of $A$-values for the limit cycles, are plotted in figures \ref{mc0},\ref{mc1},\ref{mc2}, using the numerical continuation package MATCONT \cite{dhooge}. The investigation to follow will reveal that there are also
global bifurcations which occur at interior points of some of the regions, and not only at their boundaries.

\subsubsection{$I\rightarrow II$ and $I\rightarrow III$: fold bifurcations}

\begin{figure}
	\begin{center}
		\includegraphics[width=0.42\linewidth]{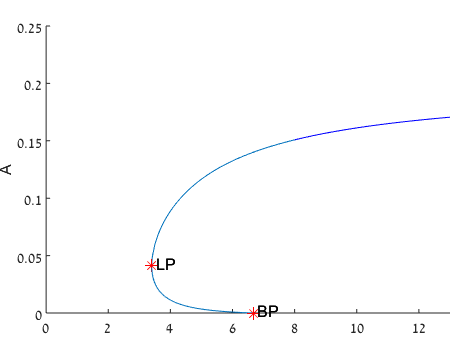}	
	\end{center}
	\caption{Left: Equilibria in dependence on $\delta$, when $p=0.15$, $\frac{\gamma}{\mu}=4$ }
	\label{mc0}
\end{figure}

When the line $\delta=4(1-p)$ (the endemicity threshold) is crossed from left to right, we have a fold (also known as limit-point) bifurcation \cite{kuznetsov}
in which the two endemic equilbria $E_1,E_2$ appear. In the case of crossing from region $I$ to region $II$ 
(that is when the point of crossing satisfies $\frac{\gamma}{\mu}<F(\delta)$, as is the case in figure \ref{mc0}), the equilibrium $E_1$ is stable 
and $E_2$ is a saddle with a one-dimensional unstable manifold - this is also known as a saddle-node bifurcation.
In the case of crossing from $I$ to $III$ (as is the case in figure \ref{mc1} and \ref{mc2}) both $E_1$ and $E_2$ are saddles, with $E_1$ having a two-dimensional unstable manifold and $E_2$ having a one-dimensional unstable manifold. 

\subsubsection{$II\rightarrow III$: transcritical bifurcation}

When we cross the invasion threshold $\delta=\frac{1}{p}$ from region $II$ to region $III$, a transcritical bifurcation 
ocurrs whereby the unstable endemic equilbirium $E_1$ merges with the stable contagion-free equilibrium $E_0$ and then disappears (its $A$ 
component becomes negative), and $E_0$ becomes unstable. 

\subsubsection{$IV\rightarrow III$: Hopf bifurcation}
\label{hopf}

\begin{figure}
	\begin{center}
		\includegraphics[width=0.42\linewidth]{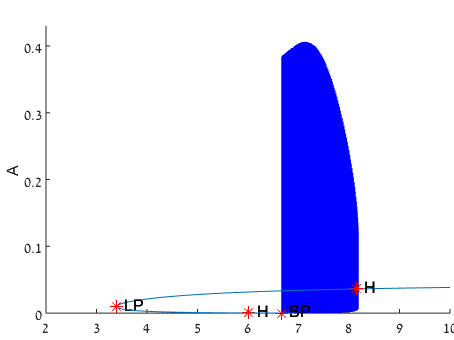}
		\includegraphics[width=0.42\linewidth]{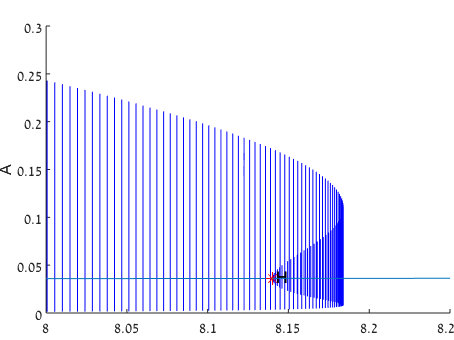}
	\end{center}
	\caption{Left: Equilibria and limit cycles in dependence on $\delta$, when $p=0.15$, $\frac{\gamma}{\mu}=20$. Right: 
		A closer look at values of the parameter $\delta$ near the Hopf bifurcation point. }
	\label{mc1}
\end{figure}

When we cross from region $III$ to region $IV$, along the curve $\frac{\gamma}{\mu}=F(\delta)$, the unique endemic equilibrium 
$E_1$ loses stability as two eigenvalues of the linearization around $E_1$ move from the left to right-hand side of the 
complex plane, leading to the birth of a limit cycle through a Hopf bifurcation \cite{kuznetsov}. For the case 
$\frac{\gamma}{\mu}=20$, this occurs at $\delta=8.14$, and the bifurcating limit cycle can be observed in figure \ref{mc1}. 
By taking a closer look at the neighborhood of the bifurcation point (figure \ref{mc1},right), we find that the bifurcation is of subcritical type: 
as $\delta$ increases beyond the critical value $\delta=8.14$, a branch of {\it{unstable}} limit cycles  is born 
out of the equilibrium point (which changes from unstable to stable). At $\delta=8.184$, this branch of limit cycles folds
back and becomes stable, and this generates the limit cycles which characterize the dynamics in region $IV$. This 
means that for $8.14<\delta<8.184$ - parameter values for which we are in region $III$, there exist both a stable endemic equilibrium
{\it{and}} a stable limit cycle, and in addition there is an unstable limit cycle - thus we have bistabilty of periodic and 
stationary behavior. It also implies that the transition from a stable endemic equilibrium to periodic behavior as $\delta$
decreases, so that we move from region $III$ to region $IV$, occurs in a discontinuous manner - when $E_1$ loses stability the stable limit cycle which characterizes the dynamics is a large one. 

\subsubsection{$IV\rightarrow V$: Saddle-node homoclinic bifurcation}

\begin{figure}
	\begin{center}
		$\delta=8\;\;\;\;\;\;\;\;\;\;\;\;\;\;\;\;\;\;\;\;\;\;\;\;\;\;\;\delta=7.5\;\;\;\;\;\;\;\;\;\;\;\;\;\;\;\;\;\;\;\;\;\;\;\;\;\;\;\delta=6.666\;\;\;\;\;\;\;\;\;$\\
		\includegraphics[width=0.3\linewidth]{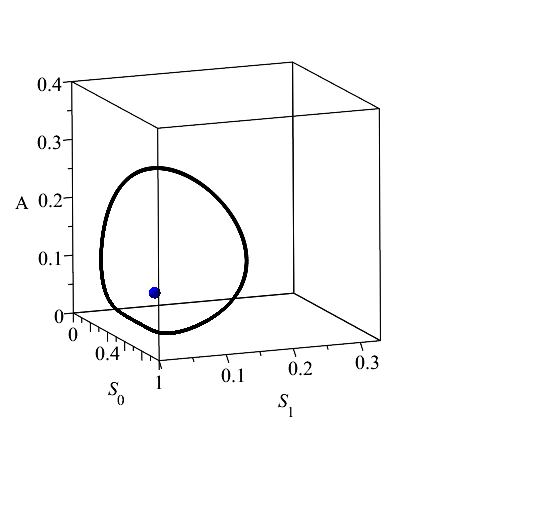}
		\includegraphics[width=0.3\linewidth]{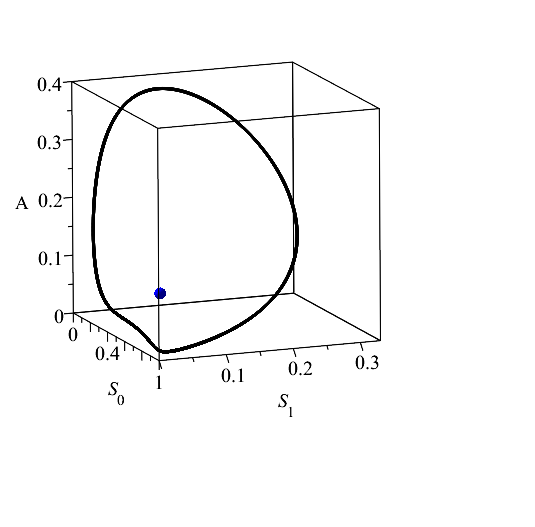}
		\includegraphics[width=0.3\linewidth]{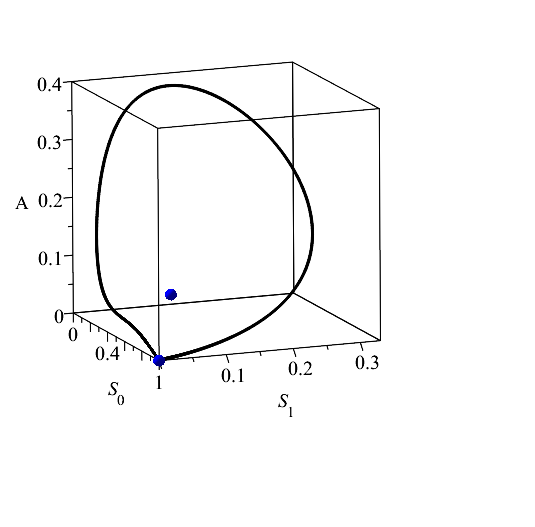}\\
		$\delta=6\;\;\;\;\;\;\;\;\;\;\;\;\;\;\;\;\;\;\;\;\;\;\;\;\;\;\;\delta=5\;\;\;\;\;\;\;\;\;\;\;\;\;\;\;\;\;\;\;\;\;\;\;\;\;\;\;\;\;\;\;\;\;\delta=4\;\;\;\;\;\;\;\;\;$\\
		\includegraphics[width=0.3\linewidth]{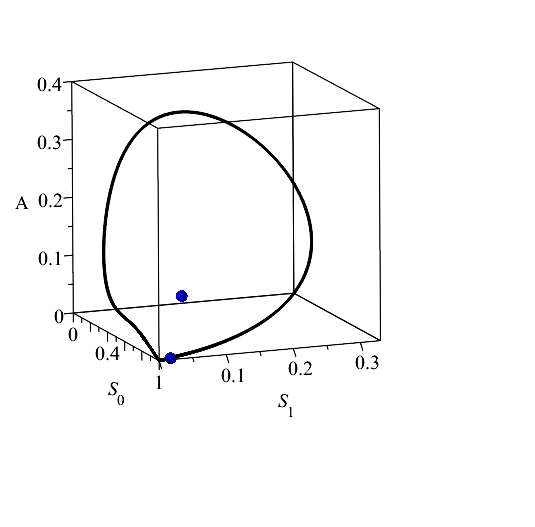}
		\includegraphics[width=0.3\linewidth]{bif_eta=20_delta=5p}
		\includegraphics[width=0.3\linewidth]{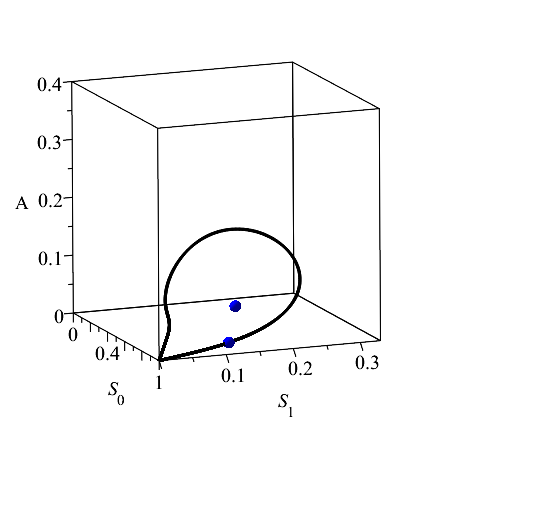}	
	\end{center}
	\caption{Saddle-node homoclinic bifurcation as $\delta$ decreases, for $p=0.15$, $\frac{\gamma}{\mu}=20$.}
	\label{sn}
\end{figure}

Referring to figure \ref{mc1} (left) we see that when $\delta$ is reduced beyond the value $\delta=\frac{1}{p}=\frac{1}{0.15}=6.666$, so that we
move from region $IV$ to region $V$, the limit cycle disappears, and we would like to understand the
type of bifurcation involved. 
We recall (section \ref{excite}) that when parameter values are in region $V$ the dynamics is characterized by excitability, stemming from the a hetroclinic loop connecting the unstable equilibrium $E_2$ 
to the stable contagion-free equilibrium $E_0$. The transition from the limit cycle to the heteroclinic loop occurs through a
saddle-node homoclinic bifurcation \cite{kuznetsov}, see figure \ref{sn}: as $\delta$ approaches $\frac{1}{0.15}$ from above, part of the limit cycle 
approaches closer to the contagion-free equilibrium $E_0$, and when $\delta=\frac{1}{0.15}$ the limit cycle touches $E_0$, thus 
forming a homoclinic (note that this implies that as $\delta$ approaches the critical value, the period of the limit cycle approaches 
infinity). As soon as $\delta<\frac{1}{0.15}$, the unstable endemic equilibrium $E_2$ bifurcates out of $E_0$, which now becomes stable, and the heteroclinic loop
$E_2\rightarrow E_0\rightarrow E_2$ is formed (note that the other unstable equilibrium $E_1$ is not involved in this bifurcation).
The limit cycle has thus been replaced by a heteroclinic loop, so that oscillations have been replaced by excitability.

\subsubsection{$V\rightarrow II$: Hopf bifurcation of an unstable limit cycle}

\begin{figure}
	\begin{center}
		\includegraphics[width=0.42\linewidth]{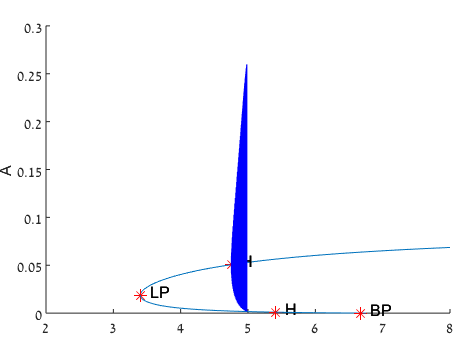}
		\includegraphics[width=0.42\linewidth]{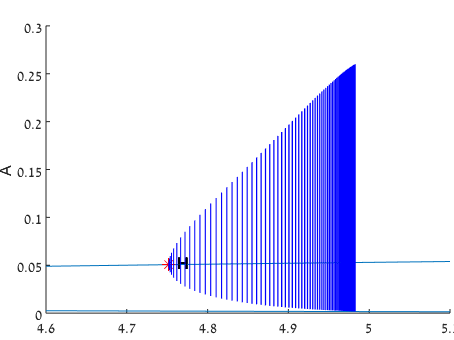}
	\end{center}
	\caption{Left: Equilibria and limit cycles in dependence on $\delta$, when $p=0.15$, $\frac{\gamma}{\mu}=10$. Right: 
		A closer look at values of the parameter $\delta$ near the Hopf bifurcation point. }
	\label{mc2}
\end{figure}

\begin{figure}
	\begin{center}
		\includegraphics[width=0.42\linewidth]{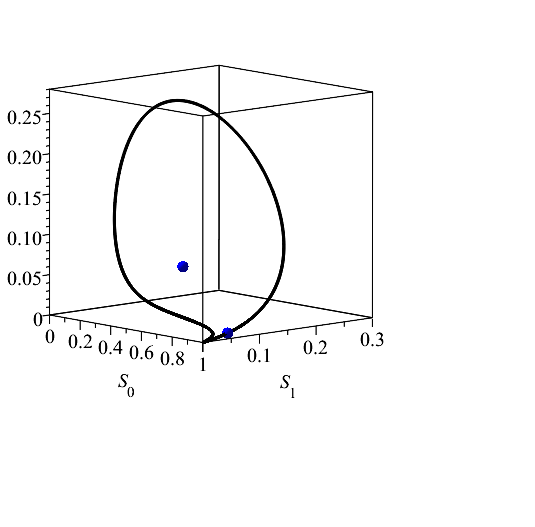}\hspace{1cm}
		\includegraphics[width=0.42\linewidth]{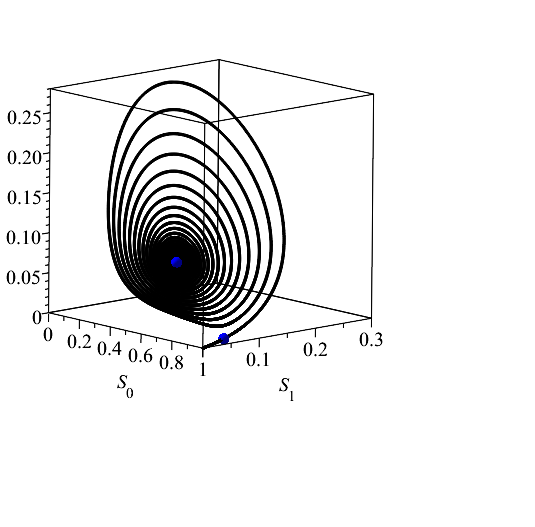}
		
	\end{center}
	\caption{For $p=0.15,\frac{\gamma}{\mu}=10$, unstable manifold of $E_2$ before ($\delta=4.9$, left) and after ($\delta=5.1$, right) the homoclinic bifurcation. }
	\label{bif}
\end{figure}

When we move from region $V$ to region $II$ across the boundary defined by the curve $\frac{\gamma}{\mu}=F(\delta)$, the endemic 
equilibrium $E_1$ becomes stable, as two eigenvalues of its linearization cross from the right to the left halves of the complex plane.
This is accompanied, as expected, by a Hopf bifurcation, in which an {\it{unstable}} limit cycle emerges from $E_1$. This can 
be seen in Figure \ref{mc2}, where we take $\frac{\gamma}{\mu}=10$ and the transition from region $V$ to region $II$
occurs at $\delta=4.752$. In this figure we also see that the unstable limit cycle which is born at $\delta=4.752$ 
disappears at $\delta=4.953$. A closer examination reveals a global bifurcation involving also the one-dimensional unstable manifold of $E_2$.
For values of $\delta$ below the bifurcation value $\delta=4.953$ (figure \ref{bif}, left), this unstable manifold forms a homoclinic loop
with $E_0$. At the bifurcation the unstable limit cycle collides with the unstable  equilibrium $E_2$,
forming a homoclinic orbit, which then vanishes, and the two parts of the unstable manifold of $E_2$ now connect to $E_0$ and to $E_1$ (figure \ref{bif}, right).

\section{Discussion}
\label{discussion}

We have shown that simple two stage contagion models with demographic turnover generate interesting nonlinear effects, which do not arise in their `classical' one-stage counterparts (SI and SIR models \cite{keeling,martcheva}). We now summarize these effects in a non-technical way, so as to highlight the qualitative conclusions that can be drawn regarding conditions under which different types of dynamics are obtained, and consider
their significance for the behavior of contagion at the population level. We expect that the broad features described below will be robust, in the sense that they will hold also under various modifications 
of the model.

(1) When the probability of adoption on first encounter is sufficiently high ($p\geq \frac{1}{2}$), the 
behavior of the two-stage contagion models is much like that of a one-stage model: there exists an invasion threshold 
$\delta=\frac{1}{p}$ such that when contagion is weak ({\it{i.e.}} the contact parameter $\delta$ satisfies $\delta\leq \frac{1}{p}$) the contagion cannot persist, while for $\delta$ above this threshold the contagion persists and approaches an endemic equilibrium. Moreover, the transition from non-contagion to contagion is continuous, in the sense that when the contact parameter is slightly above 
the threshold, the extent of contagion ($A$) at equilibrium will be small.

When the probability of adoption on first encounter is sufficiently low ($p<\frac{1}{2}$), new qualitative features emerge.
In this case the dynamical behavior of the model depends on two factors: the contact parameter $\delta$, and the duration of the adoption period relative to the average residence time of an individual in the population. The different phenomena which occur, depending on these 
two factors, are summarized below.

(2) When the contact paramater is below the endemicity threshold ($\delta<4(1-p)$), contagion will not spread in the population.

(3) When the contact parameter is above the invasion threshold ($\delta>\frac{1}{p}$), we observe different 
behaviors, depending on the duration of adoption: 

(i) If the mean duration of adoption ($\gamma^{-1}$) is sufficiently long relative to the mean 
residence time of individuals in the population ($\mu^{-1}$) -- including the extreme case $\gamma=0$ which corresponds to the 
model with permanent adoption, the contagion becomes established at a constant level (a stable endemic equilibrium), regardless of the initial number of adopters. 

(ii) If the mean duration of adoption is sufficiently short (that is $\frac{\gamma}{\mu}$ is sufficiently large), sustained
periodic oscillations occur - corresponding to cyclic epidemics of contagion. 
These oscillations are an emergent phenomenon at the population level, arising from
interactions among individuals, each of which displays no cyclic behavior - recall that in this model each individual who abandons the innovation does not re-adopt it.  We therefore have a mechanism for the endogenous generation of periodic fads and fashions. We note that a quite different mechanism capable of generating periodic fashions, based on `snobs' and `followers', is modelled in \cite{apriasz,callahan}. 
We note that endogenous oscillations do not occur in the basic models of mathematical epidemiology, and some 
special mechanisms are required to generate such oscillations, the most prominent being temporary immunity with delay \cite{hethcote2,yuan}.
Our results show that two-stage contagion is another mechanism which induces periodic oscillations, and it 
appears that it is among the simplest mechanisms producing this effect.

(iii) In a narrow intermediate range of values of the mean duration of adoption, endemic equilibrium and periodic oscillations
are both stable, so that contagion will be maintained either at a constant level or in the form of cyclic epidemics, depending on initial conditions.

(4) When the contact parameter is above the endemicity threshold but below the invasion threshold ($4(1-p)<\delta<\frac{1}{p}$), we observe different behaviors, depending on the duration of adoption: 

(i) If the mean duration of adoption ($\gamma^{-1}$) is sufficiently long relative to the mean 
residence time of individuals in the population ($\mu^{-1}$) (including in the limit $\gamma=0$, corresponding to permanent adoption),
we have bistability of the contagion-free and endemic equilibria, (`alternative stable states', \cite{scheffer}),  giving rise to 
a `critical mass' threshold, 
so that contagion will only `catch' if sufficiently many individuals adopt it at the start. An important implication of this bistability is that eradication of an established contagion will require reducing the contact parameter to a much lower value than the invasion threshold. For example, if $p=0.1$, then the invasion threshold is $\delta=\frac{1}{p}=10$, but eradicating an existing endemic contagion will require reducing $\delta$ below $\delta=4(1-p)=3.6$. Related to this is the discontinuous transition which occurs under a varying contact parameter: when the contagion-free state is de-stabilized as 
$\delta$ crosses the invasion threshold from below, a jump from no contagion to a high level of contagion will occur. Similarly, 
if a cotagion is already established and the contact parameter is reduced until it reaches the endemicity threshold, a large contagion will
disappear without warning.
This type of `critical transition' or `regime shift' phenomenon \cite{scheffer} can provide an explanation for rapid opinion shifts and dramatic behavioral changes which can arise under minor changes in external conditions \cite{kuran,van}. Under this explanation, the discontinuous transition is a collective effect arising from the interactions among individuals - the behavior of individuals changes only in a gradual way as the contact parameter varies.

The bistability and hysteresis effects described above do not occur in the basic `one stage' epidemiological models, in which transition from 
the contagion-free state to endemicity is continuous. However such effects do occur in some more elaborate models,
and are known under the term `backward bifurcation'. Various epidemiological mechanisms are known to induce 
backward bifurcations, e.g. exogenous reinfection of latently infected individuals, imperfect vaccination, and risk structure 
\cite{gumel}. Our results show that contagion with stages is another mechanism which generates backward bifurcation.

(ii) If the mean duration of adoption is sufficently short (that is $\frac{\gamma}{\mu}$ is sufficently large),
then contagion cannot become endemic, but we observe the phenomenon of
excitability:  starting with a small fraction of initial
adopters, a large epidemic develops before the contagion fades. This contrasts with one-stage contagion models, in which, below the invasion threshold, the number of adopters will always decrease, whatever its initial value. Thus the two-stage model can account for 
large contagion epidemics which do not become endemic, despite the renewal of the susceptible population provided by the 
demographic turover.

Developing the mathematical theory of social contagion requires classifying relevant mechanisms at the micro-level, and exploring 
their dynamical consequences at the population level using mathematical modelling. Simple models, like the one considered here,
show that a combination of basic mechanisms (here, two-stage contagion and demographic turnover) can give rise to rich phenomena, suggestive of some of the complexities found in social systems. As always, the fact that a mathematical model can produce a phenomenon 
which is reminiscent of a real-world one is far from proof that the mechanisms described by the model are those 
responsible for the real effect, but it does constitute a proof-of-principle that the mechanisms involved are capable of 
producing the effect \cite{kokko}. It would be of great interest to attempt a direct validation of a two-stage (or multi-stage)
contagion model by fitting it to empirical data.

\bibliography{contagiont}{}
\bibliographystyle{plain}
\end{document}